\documentclass[12pt,tightenlines]{revtex4}
\usepackage{amsfonts}
\usepackage{amsmath}
\usepackage{txfonts}
\usepackage{graphicx}
\usepackage{amssymb}
\usepackage{color}
\usepackage{wrapfig,lipsum,booktabs}
\usepackage{floatrow}
\usepackage{epstopdf}
\usepackage{hyperref}
\usepackage[normalem]{ulem}

\newcommand{\beginsupplement}{
        \setcounter{table}{0}
        \renewcommand{\thetable}{S\arabic{table}}
        \setcounter{figure}{0}
        \renewcommand{\thefigure}{S\arabic{figure}}
     }
     
\begin{document}
\title{Transient cell assembly networks encode persistent spatial memories}
\author{Andrey Babichev$^{1,2}$ and Yuri Dabaghian$^{1,2*}$}
\affiliation{$^1$Jan and Dan Duncan Neurological Research Institute, Baylor College of Medicine,
Houston, TX 77030, \\
$^2$Department of Computational and Applied Mathematics, 
Rice University, Houston, TX 77005\\
$^{*}$e-mail: dabaghia@bcm.edu}
\vspace{17 mm}
\date{\today}
\vspace{17 mm}
\begin{abstract}
While cognitive representations of an environment can last for days and even months, the synaptic architecture 
of the neuronal networks that underlie these representations constantly changes due to various forms of synaptic 
and structural plasticity at a much faster timescale. This raises an immediate question: how can a transient network 
maintain a stable representation of space? In the following, we propose a computational model for describing 
emergence of the hippocampal cognitive map ina network of transient place cell assemblies and demonstrate, using 
methods of algebraic topology, that such a network can maintain a robust map of the environment.
\end{abstract}
\maketitle

\newpage

\section{Introduction}
\label{section:intro}
The mammalian hippocampus plays a major role in spatial cognition, spatial learning and spatial memory by producing 
an internalized representation of space---a cognitive map of the environment \cite{Tolman, OKeefe,Nadel,McNaughton}. 
Several key observations shed light on the neuronal computations responsible for implementing such a map. The first 
observation is that the spiking activity of the principal cells in the hippocampus is spatially tuned. In rats, these neurons, 
called place cells, fire only in restricted locations---their respective place fields \cite{Best1}. As demonstrated in many 
studies, this simple principle allows decoding the animal's ongoing trajectory \cite{Brown1,Guger}, its past navigational 
experience \cite{Carr}, and even its future planned routs \cite{Pfeiffer,Dragoi1,Dragoi2} from the place cell's spiking activity. 

The second observation is that the spatial layout of the place fields---the place field map---is ``flexible'': as the environment 
is deformed, the place fields shift and change their shapes, while preserving most of their mutual overlaps, adjacency and 
containment relationships \cite{Gothard,Leutgeb,Wills,Touretzky}. Thus, the sequential order of place cells' (co)activity induced 
by the animal's moves through morphing environment remains invariant within a certain range of geometric transformations
\cite{Diba,eLife,Alvernhe,Poucet,Wu}. This implies that the place cells' spiking encodes a coarse framework of qualitative 
spatiotemporal relationships, i.e., that the hippocampal map is topological in nature, more similar to a schematic subway map 
than to a topographical city map \cite{eLife}.

The third observation concerns the synaptic architecture of the (para)hippocampal network: it is believed that groups of 
place cells that demonstrate repetitive coactivity form functionally interconnected ``assemblies,'' which together drive their 
respective ``reader-classifier'' or ``readout'' neurons in the downstream networks \cite{Harris1,Buzsaki1}. The activity of 
a readout neuron actualizes the qualitative relationships between the regions encoded by the individual place cells, thus 
defining the type of spatial connectivity information encoded in the hippocampal map \cite{Schemas}.

A given cell assembly network architecture appears as a result of spatial learning, i.e., it emerges from place cell 
coactivities produced during an animal's navigation through a particular place field map, via a ``fire-together-wire-together'' 
plasticity mechanism \cite{Caroni,Chklovskii}. However, a principal property of the cell assemblies is that they may not only
form, but also or disband as a result of a depression of synapses caused by reduction or cessation of spiking activity over a 
sufficiently long timespan \cite{Wang}. Some of the disbanded cell assemblies may later reappear during a subsequent period 
of coactivity, then disappear again, and so forth. Electrophysiological studies suggest that the lifetime of the cell assemblies 
ranges between minutes \cite{Kuhl,Murre} to hundreds of milliseconds \cite{Atallah,Bartos,Mann,Whittington,Bi}. In contrast, 
spatial memories in rats can last much longer \cite{Meck,Clayton,Brown2}, raising the question: how can a large-scale spatial 
representation of the environment be stable if the neuronal stratum that computes this representation changes on a 
much faster timescale? 

The hypothesis that the hippocampus encodes a topological map of the environment allows this question to be addressed 
computationally. Below, we propose a phenomenological model of a transient hippocampal network and use methods of 
algebraic topology to demonstrate that a large-scale topological representation of the environment encoded by this network 
may remain stable despite the transience of neuronal connections.

\section{The Model}
\label{section:model}

\begin{figure} 
\includegraphics[scale=0.84]{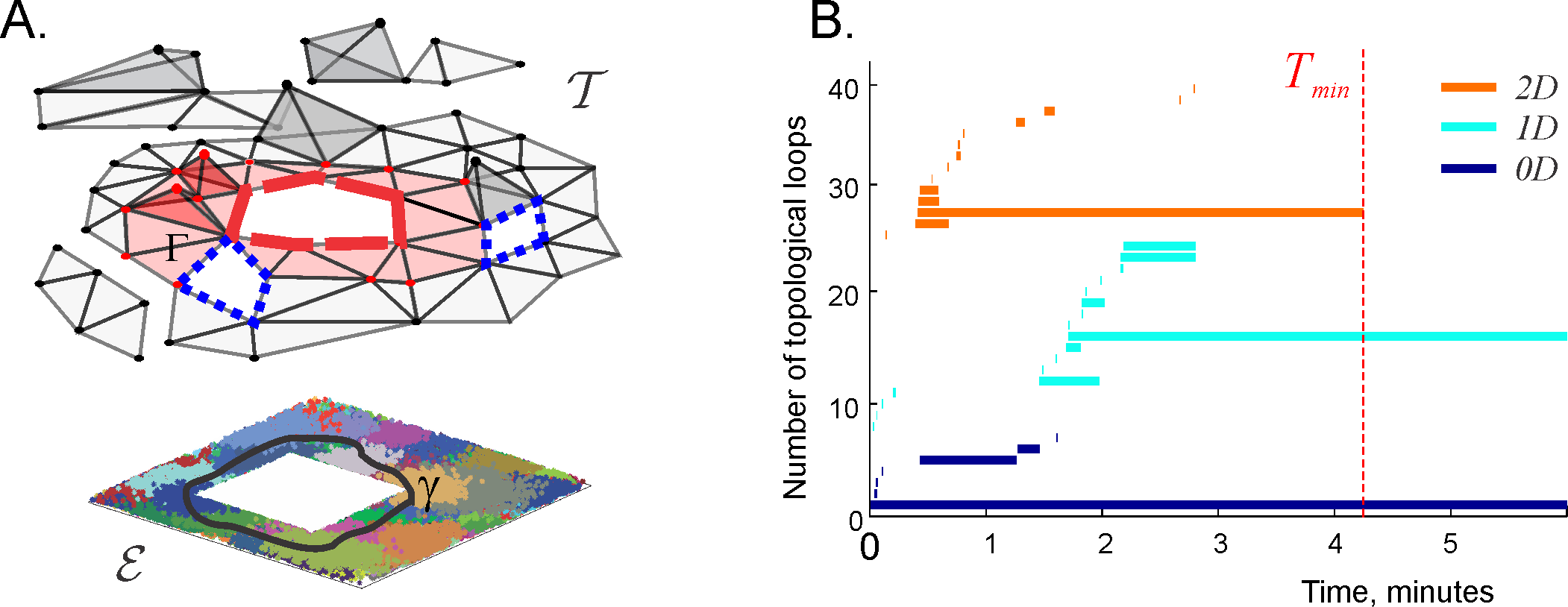}
\caption{\label{Figure1} 
{\footnotesize\textbf{Schematic representation of place cell coactivity in the cell assembly complex $\mathcal{T}$}. 
(\textbf{A}) Bottom: Simulated place map covering a small planar environment $\mathcal{E}$ that contains a square hole 
in the middle. The clusters of colored dots represent the place fields, and a typical path traversed by the animal is shown 
by the black loop $\gamma$. The coactivity complex $\mathcal{T}$ induced from the coactivity of the place cells (on the top) 
provides a topological representation of environment. The hole in the middle of $\mathcal{T}$ (red dashed line) corresponds 
to the central hole in the environment $\mathcal{E}$. The non-contractible closed chain of simplexes---a simplicial path 
$\Gamma$---corresponds to the non-contractible physical path $\gamma$. In general, the complex can be fragmented in 
pieces and contain gaps (holes encircled by blue loops), which represent topological noise, rather than information about 
the environment. (\textbf{B}) Timelines of zero-dimensional ($0D$), one-dimensional ($1D$) and two-dimensional ($2D$) 
topological loops encoded in a coactivity complex $\mathcal{T}$ which represents the environment $\mathcal{E}$. For as 
long as a given $0D$ loop persists, it indicates that the coactivity complex contains the corresponding connectivity component. 
A persisting $1D$ loop represents a noncontractible hole in $\mathcal{T}$ (see red hole in $\mathcal{T}$ on the left panel). 
A persistent $2D$ loop represents a noncontractible $2D$ sphere in $\mathcal{T}$. 
In the illustrated case, the $0D$ spurious loops disappear in about 1.5 minutes, indicating that the coactivity complex 
$\mathcal{T}$ fuses into one piece at that time. The $1D$ spurious disappear when all the spurious holes in 
$\mathcal{T}$ close up in about 2.8 min, and the $2D$ loops disappear by $T_{\min} = 4.2$ min (dashed vertical line), 
at which point the topology of $\mathcal{T}$ becomes equivalent to the topology of the environment.}}
\end{figure}

In \cite{PLoS,Arai}, we proposed a computational approach to integrating the information provided by the individual place 
cells into a large-scale topological representation of the environment, based on several remarkable parallels between 
the elements of hippocampal physiology and certain notions of algebraic topology. There, we regarded a particular 
collection of coactive place cells $c_{1}$, $c_{2}$, ..., $c_n$ as an abstract ``coactivity simplex'' $\sigma = [c_{1}, 
c_{2}, ...,c_{n}]$, which may be visualized (for $n \leq 4$) or apprehended (for $n\geq 5$) as an $(n-1)$-dimensional 
polyhedron \cite{Aleksandrov}. As a result, the pool of coactivities is represented by a simplicial ``coactivity'' complex 
$\mathcal{T}$, which provides a link between the cellular and the net systemic level of the information processing. 
Just like simplexes, the individual cell groups provide local information about the environment, but together, as a neuronal 
ensemble, they represent space as whole---as the simplicial complex. Numerical experiments \cite{PLoS,Arai,Curto} 
backed up by a remarkable theorem due to P. Alexandrov \cite{Alexandroff} and E. \v{C}ech's \cite{Cech} point out 
that $\mathcal{T}$ correctly represents the topological structure of the rat's environment and may serve as a schematic 
model of the hippocampal map \cite{Schemas}. For example, the paths traveled by the rat are represented by the 
``simplicial paths''---chains of simplexes in $\mathcal{T}$ that capture certain qualitative properties of their physical 
counterparts (see \cite{Dabaghian1,Novikov} and Fig.~\ref{Figure1}A).

Of course, producing a faithful representation of the environment from place cell coactivity requires learning. In the 
model, this process is represented by the dynamics of the coactivity complex's formation. At every moment of time, 
the coactivity complex represents only those place cell combinations that have exhibited (co)activity. As the animal 
begins to explore the environment, the newly emerging coactivity complex is small, fragmented and contains many 
holes, most of which do not correspond to physical obstacles or to the regions that have not yet been visited by the 
animal. These ``spurious'' structures tend to disappear as the pool of place cell coactivities accumulates. Numerical 
simulations show that, if place cells operate within biological parameters \cite{PLoS}, the topological structure of 
$\mathcal{T}$ becomes equivalent to the topological stricture of the environment within minutes. The minimal time 
$T_{\min}$ required to produce a correct topological representation of the environment can be used as an estimate 
for the time required to learn spatial connectivity (Fig.~\ref{Figure1}B, \cite{PLoS,Arai,Curto}).

\begin{figure} 
\includegraphics[scale=0.84]{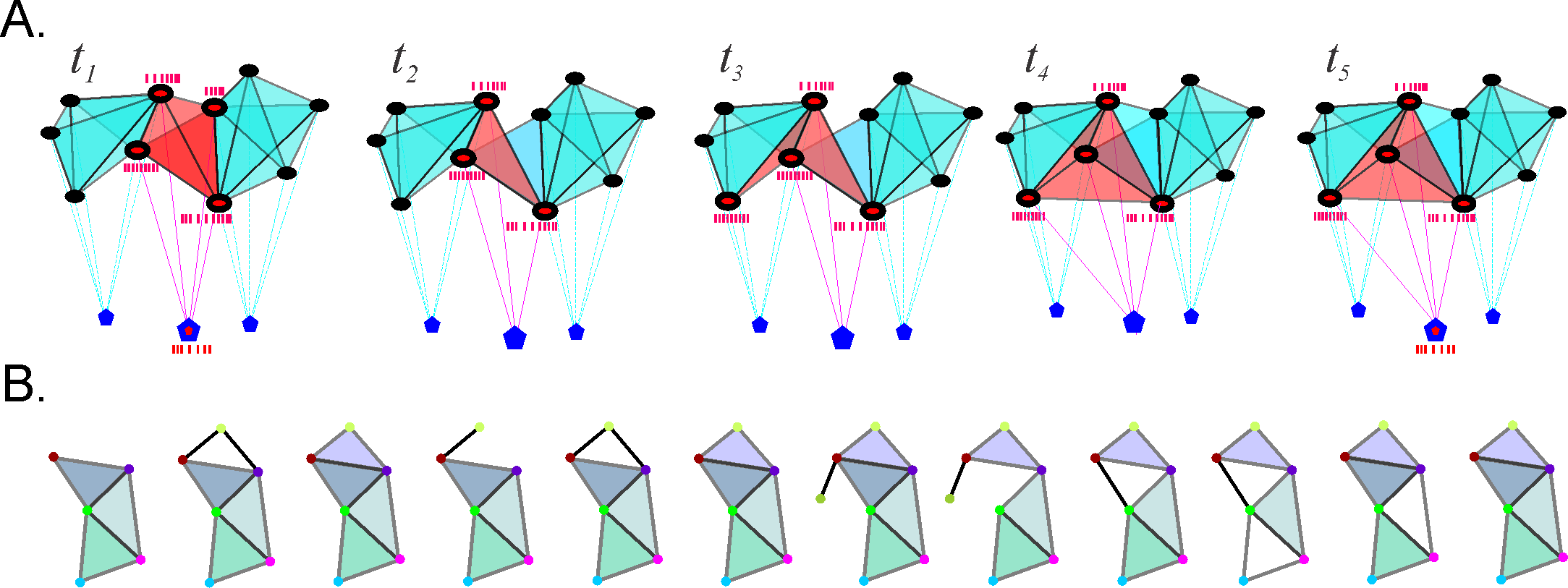}
\caption{\label{Figure2} {\footnotesize
\textbf{Place cell assemblies and flickering coactivity complexes}. (\textbf{A}) Functionally 
interconnected groups of place cells (place cell assemblies) are schematically represented by fully interconnected 
cliques. The place cells (small disks) in a given assembly $\varsigma$ are synaptically connected to the corresponding 
readout neuron $n_{\varsigma}$ (pentagons below). An assembly $\varsigma$ ignites (red clique/tetrahedron in the middle) 
when its place cells elicit jointly a spiking response from the readout neuron $n_{\varsigma}$ (active cells have 
red centers). A cell assembly may be active at a certain moment of time, then disactivate, then become active again, 
and so forth. If a certain cell assembly ceases to ignite and another combination of place cells begins to exhibit frequent 
coactivity, the old cell assembly is replaced by new one. (\textbf{B}) The formation and disbanding of the cell assemblies 
is schematically represented in the ``flickering'' coactivity complex, in which the maximal simplexes appear and disappear, 
representing appearance and disappearance of the cell assemblies in the hippocampal network.} }
\end{figure}

\textbf{Coactivity complexes}. A specific algorithm used to implement a coactivity complex may be designed to 
incorporate physiological information about place cell cofiring at different levels of detail. In the simplest case, every 
observed group of coactive place cells contributes a simplex; the resulting coactivity complex $\mathcal{T}$ (referred 
to as the \v{C}ech coactivity complex in \cite{PLoS,Arai}) makes no reference to the structure of the hippocampal 
network or to the cell assemblies, and gives a purely phenomenological description of the information contained in 
the place cell coactivity. In a more detailed approach, the maximal simplexes of the coactivity complex (i.e., the 
simplexes that are not subsimplexes of any larger simplex) may be selected to represent ignitions of the place cell 
assemblies, rather than arbitrary place cell combinations. The combinatorial arrangement of the maximal simplexes 
in the resulting ``cell assembly coactivity complex,'' denoted $\mathcal{T}_{CA}$, schematically represents the 
network of interconnected cell assemblies \cite{Schemas,Babichev1} (Fig.~\ref{Figure2}A).

The specific algorithm of constructing the complex $\mathcal{T}_{CA}$ may also reflect how neuronal coactivity 
is processed by the readout neurons. If these neurons function as ``coincidence detectors,'' i.e., if they react to the 
spikes received within a short coactivity \emph{detection} period $w$ (typically, $w \approx 200-250$ milliseconds 
\cite{Arai, Mizuseki}), then the maximal simplexes in the corresponding coincidence detection coactivity complex 
(denoted $\mathcal{T}_{\ast}$) will appear instantaneously at the moments of the cell assemblies' ignitions 
\cite{Babichev1,Hoffmann}. Alternatively, if the readout neurons integrate the coactivity inputs from smaller parts 
of their respective assemblies over an extended coactivity \emph{integration} period $\varpi$ \cite{Konig,Ratte}, then 
the appearance of the maximal simplexes in the corresponding input integration coactivity complex (denoted as 
$\mathcal{T}_{+})$ will extend over time, reflecting the dynamics of synaptic integration. The time course of the 
maximal simplexes' appearance affects the rate at which large-scale topological information is accumulated and 
hence controls the model's description of spatial learning.

\textbf{Computational implementation} of the coactivity complexes $\mathcal{T}_{\ast}$ and $\mathcal{T}_{+}$ 
is as follows. The maximal simplexes of the coincidence detection coactivity complex $\mathcal{T}_{\ast}$ are 
selected from the pool of the most frequently appearing groups of simultaneously coactive place cells \cite{Babichev1}. 
To model an input integrator coactivity complex $\mathcal{T}_{+}$, we first built a connectivity graph $G$ that 
represents pairwise place cell coactivities observed within a certain period $\varpi$. Then we build the associated 
clique complex $\mathcal{T}_{\varsigma}(G)$, i.e., we view the maximal, fully interconnected subgraphs of the 
graph $G$, which are its cliques $\varsigma$, as simplexes of $\mathcal{T}_{\varsigma}$ (for details see Methods 
in \cite{Babichev1} and \cite{Jonsson}). The process of assembling the fully interconnected cliques from pairwise 
connections is designed to model the process of integrating the spiking inputs in the cell assemblies, so that the 
resulting clique coactivity complex $\mathcal{T}_{\varsigma}$ serves as a model of the input integration cell 
assembly network.

Numerical simulations show that, for a given population of place cells, the clique complex $\mathcal{T}_{\varsigma}$ 
is typically larger and forms faster than the coincidence detector (\v{C}ech) complex $\mathcal{T}_{\varsigma}$, 
and, as a result, $\mathcal{T}_{\varsigma}$ reproduces the topological structure of the environment more reliably 
\cite{Schemas,Babichev1}. Moreover, the coincidence detection coactivity complexes can be viewed as a specific 
case of the input integration coactivity complexes: as the integration period shrinks and approaches the coactivity 
period $\varpi \to w$, the input integration coactivity complex $\mathcal{T}_{\varsigma}$ reduces to the coincidence 
complex $\mathcal{T}_{\varsigma}$. For these reasons, in the following we will model only the input integration, 
i.e., clique coactivity complexes.

\textbf{Instability of the cell assemblies}. In our previous work \cite{Babichev1}, the frequencies of the cell assemblies' 
appearances, $f_{\varsigma}$, were computed across the entire navigation period $T_{tot}$, i.e., the cell assemblies 
were presumed to exist from the moment of their first appearance for as long as the navigation continued. In order 
to model cell assemblies with finite lifetimes, these frequencies should be evaluated within shorter periods 
$\varpi_{\varsigma} < T_{tot}$. Physiologically, $\varpi_{\varsigma}$ can be viewed as the period during which the 
readout neuron $n_{\varsigma}$ may connect synaptically to a particular combination of coactive place cells, i.e., form 
a cell assembly $\varsigma$, retain these connections, and respond to subsequent ignitions of $\varsigma$. In a 
population of cell assemblies, the integration periods can be distributed with a certain mode $\varpi$ and a variance 
$\Delta_{\varpi}$. However, in order to simplify the approach, we will make two assumptions. First, we will describe 
the entire population of the readout neurons in terms of the integration period of a typical readout neuron, i.e., describe 
the ensemble of readout neurons with a single parameter, $\varpi$. Second, we will assume that the integration periods 
of all neurons are synchronized, i.e., that there exists a globally defined coactivity integration window of width $\varpi$ 
during which the entire population of the readout neurons synchronously processes coactivity inputs from their respective 
place cell assemblies. In such case, $\varpi$ can be viewed as a period during which the cell assembly network processes 
the ongoing place cell spiking activity. Below we demonstrate that these restrictions result in a simple model that allows 
describing a population of finite lifetime cell assemblies and show that the resulting cell assembly network, for a sufficiently 
large $\varpi$, reliably encodes the topological connectivity of the environment.

\textbf{Computational model of the transient cell assembly network}. A network of rewiring cell assemblies is represented 
by a coactivity complex with fluctuating or ``flickering'' maximal simplexes. To build such a complex, denoted $\mathcal{F}$, 
we implement a ``sliding coactivity integration window'' approach. First, we identify the maximal simplexes that emerge 
within the first $\varpi$-period after the onset of the navigation $\varpi_{1}$ based on the place cell activity rates evaluated 
within that window, $f_{\varsigma}(\varpi_{1})$, and construct the corresponding input integration coactivity complex 
$\mathcal{F}(\varpi_{1})$. Then the algorithm is repeated for the subsequent windows $\varpi_{2}$, $\varpi_{3}$,... which 
are obtained by shifting the starting window $\varpi_{1}$ over small time steps $\Delta t$. Since consecutive windows overlap, 
the corresponding coactivity complexes $\mathcal{F}(\varpi_{2})$, $\mathcal{F}(\varpi_{2})$, ... consist of overlapping 
sets of maximal simplexes. A given maximal simplex $\varsigma$ (defined by the set of its vertexes) may appear in a chain 
of consecutive windows $\varpi_{1}$, $\varpi_{2}$, ..., $\varpi_{k-1}$ then disappear at a step $\varpi_{k}$ (i.e., 
$\varsigma \in \mathcal{F}(\varpi_{k-1}$), but $\varsigma \notin \mathcal{F}(\varpi_{k})$), then reappear in a later 
window $\varpi_{l}$, then disappear again, and so forth (Fig.~\ref{Figure2}). The midpoint $t_{k}$ of the window in 
which the maximal simplex $\varsigma$ has (re)appeared defines the moment of $\varsigma$'s (re)birth, and the midpoints 
of the windows were is disappears, are viewed as the times of its deaths. Indeed, one may use the left or the right end 
of the shifting integration window, which would affect the endpoints of the navigation, but not the net results discussed 
below. As a result, the lifetime $\delta t_{\varsigma,k}$ of a cell assembly $\varsigma$ between its $k$-th consecutive 
appearance and disappearance can be as short as $\Delta t$ (if $\varsigma$ appears within $\varpi_{k}$ and disappears 
at the next step, within $\varpi_{k+1}$, or as long as $T_{tot}$ - $\varpi$ in the case that $\varsigma$ appears at the 
first step and never disappears. However, a typical maximal simplex exhibits a spread of lifetimes that can be characterized 
by a half-life, as we will discuss below. 

It is natural to view the coactivity complexes $\mathcal{F}(\varpi_{i})$ as instances of a single \emph{flickering coactivity 
complex} $\mathcal{F}_{\varpi}$, $\mathcal{F}_{\varpi}(t_{i}) =\mathcal{F}(\varpi_i)$, having appearing and disappearing 
maximal simplexes (see Fig.~\ref{Figure2}B and \cite{Babichev2}). In the following, we will use $\mathcal{F}_{\varpi}$ 
as a model of transient cell assembly network and study whether such a network can encode a stable topological map of 
the environment on the moment-by-moment basis.

\begin{figure} 
\includegraphics[scale=0.84]{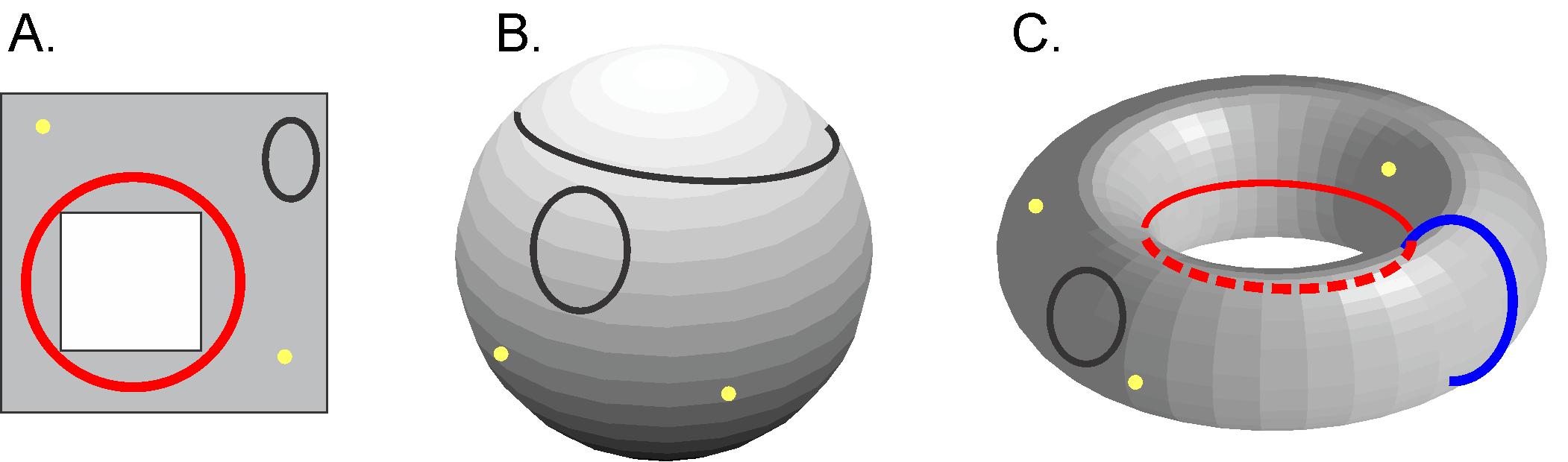}
\caption{\label{Figure3} {\footnotesize \textbf{Topological shapes defined in terms of topological loops}. (\textbf{A}) Any two 
zero-dimensional ($0D$) loops (i.e., points, yellow dots) in the two-dimensional space $\mathcal{E}$ navigated by the 
simulated rat (see Fig.~\ref{Figure1}A) can be matched with one another via continuous moves. This implies that all 
$0D$ loops are topologically equivalent to a single ``representative'' loop, i.e., that zeroth Betti number of $\mathcal{E}$ 
is $b_{0}(\mathcal{E}) = 1$. The one-dimensional ($1D$) loops are of two types: some contract to a point (e.g., the 
black loop in the corner) and others (e.g., the red loop in the middle) are non-contractible, signaling the existence of a 
topological obstruction---the central hole, which is the main topological feature of $\mathcal{E}$. Thus, $b_{1}(\mathcal{E}) = 1$.
Since the entire space $\mathcal{E}$ can be contracted to the $1D$ rim of the central hole, there are no higher dimensional 
noncontractible topological loops in $\mathcal{E}$, $b_{n > 1}(\mathcal{E}) = 0$. The net barcode of $\mathcal{E}$ 
is therefore $\mathfrak{b}(\mathcal{E}) = (1,1,0,0,...)$. (\textbf{B}) On a two-dimensional ($2D$) sphere $S^{2}$, 
every $1D$ loop can be contracted to a point (hence $b_{1}(S^{2}) = 0$), and all points can be transformed along the 
surface of $S^{2}$ into a single representative $0D$ loop (hence $b_{0}(S^{2}) = 1$). By itself, the sphere is a $2D$ 
loop (hence $b_{2}(S^{2}) = 1$), and since the sphere does not extend into higher dimensions, the rest of the Betti 
numbers vanish, $b_{n >2}(S^{2}) = 0$. Thus, the topological barcode of a sphere is $\mathfrak{b}(S^{2}) = (1,0,1,0,0,...)$. 
(\textbf{C}) A two-dimensional torus $T^{2}$ contains two inequivalent types of noncontractible $1D$ cycles, represented 
by the red and the blue loops, implying that $b_{1}(T^{2}) = 2$. The other Betti numbers in the $T^{2}$ case are the 
same as in the $S^{2}$ case, $b_{0}(T^{2}) = 1$, $b_{2}(T^{2}) = 1$ and $b_{n>2}(T^{2}) = 0$. Thus, the topological 
barcode of $T^{2}$ is $\mathfrak{b}(T^{2}) = (1,2,1,0,0,...)$.}}
\end{figure} 

\textbf{The large-scale topology}. The topological structure of a space $X$ can be described in terms of the topological 
loops that it contains, i.e., in terms of its non-contractible surfaces counted up to topological equivalence. A more basic 
topological description of $X$ is provided by simply counting the topological loops in different dimensions, i.e., by specifying 
its Betti numbers $b_n(X)$ \cite{Hatcher}. The list of the Betti numbers of a space $X$ is known as its topological barcode, 
$\mathfrak{b} (X) = (b_{0}(X), b_{1}(X), b_{2}(X), ...)$, which in many cases captures the topological identity of topological 
spaces \cite{Ghrist}. For example, the environment $\mathcal{E}$ shown at the bottom of Fig.~\ref{Figure1}A has the 
topological barcode $\mathfrak{b} (\mathcal{E}) = (1,1,0,... )$, which implies that $\mathcal{E}$ is topologically 
(homotopically) equivalent to an annulus (Fig.~\ref{Figure3}A). Other familiar examples of topological shapes identifiable 
via their topological barcodes are a two-dimensional sphere $S^{2}$ and the torus $T^{2}$ with the barcodes 
$\mathfrak{b} (S^{2}) = (1,0,1,0,... )$ and $\mathfrak{b} (T^{2}) = (1,2,1,0,... )$ respectively (Fig.~\ref{Figure3}B,C). 
For the mathematically oriented reader, we note that the matching of topological barcodes does not always imply topological 
equivalence but, in the context of this study, we disregard effects related to torsion and other topological subtleties.

In the following, we compute the topological barcode of the flickering coactivity complex at each moment of time, 
$\mathfrak{b}(\mathcal{F}_{\varpi}(t_{i}))$, and compare it to the topological barcode of the environment, 
$\mathfrak{b}(\mathcal{E})$. If, at a certain moment $t_{i}$, these barcodes do not match, the coactivity complex 
$\mathcal{F}_{\varpi}(t_{i})$ and $\mathcal{E}$ are topologically distinct, i.e., the coactivity complex 
$\mathcal{F}_{\varpi}$ misrepresents $\mathcal{E}$ at that particular moment. In contrast, if the barcode of 
$\mathcal{F}_{\varpi}(t_{i})$ is ``physical,'' i.e., coincides with $\mathfrak{b}(\mathcal{E})$, then the coactivity 
complex provides a faithful representation of the environment. More conservatively, one may compare only the 
physical dimensions of the barcodes $\mathfrak{b}(\mathcal{F}_{\varpi})$ and $\mathfrak{b}(\mathcal{E})$, 
i.e., $0D$, $1D$, $2D$ loops, or the dimensions containing the nontrivial $0D$ and $1D$ loops for the environment 
shown on Fig.~\ref{Figure1}A. Using the methods of persistent homology 
\cite{Ghrist, Zomorodian,Edelsbrunner}, we compute the minimal time required to produce the correct topological 
barcode within every integration window, which allows us to describe the rate at which the topological information 
flows through the simulated hippocampal network and discuss biological implications of the results. 

\section{Results}
\label{section:results}

\begin{figure} 
\includegraphics[scale=0.84]{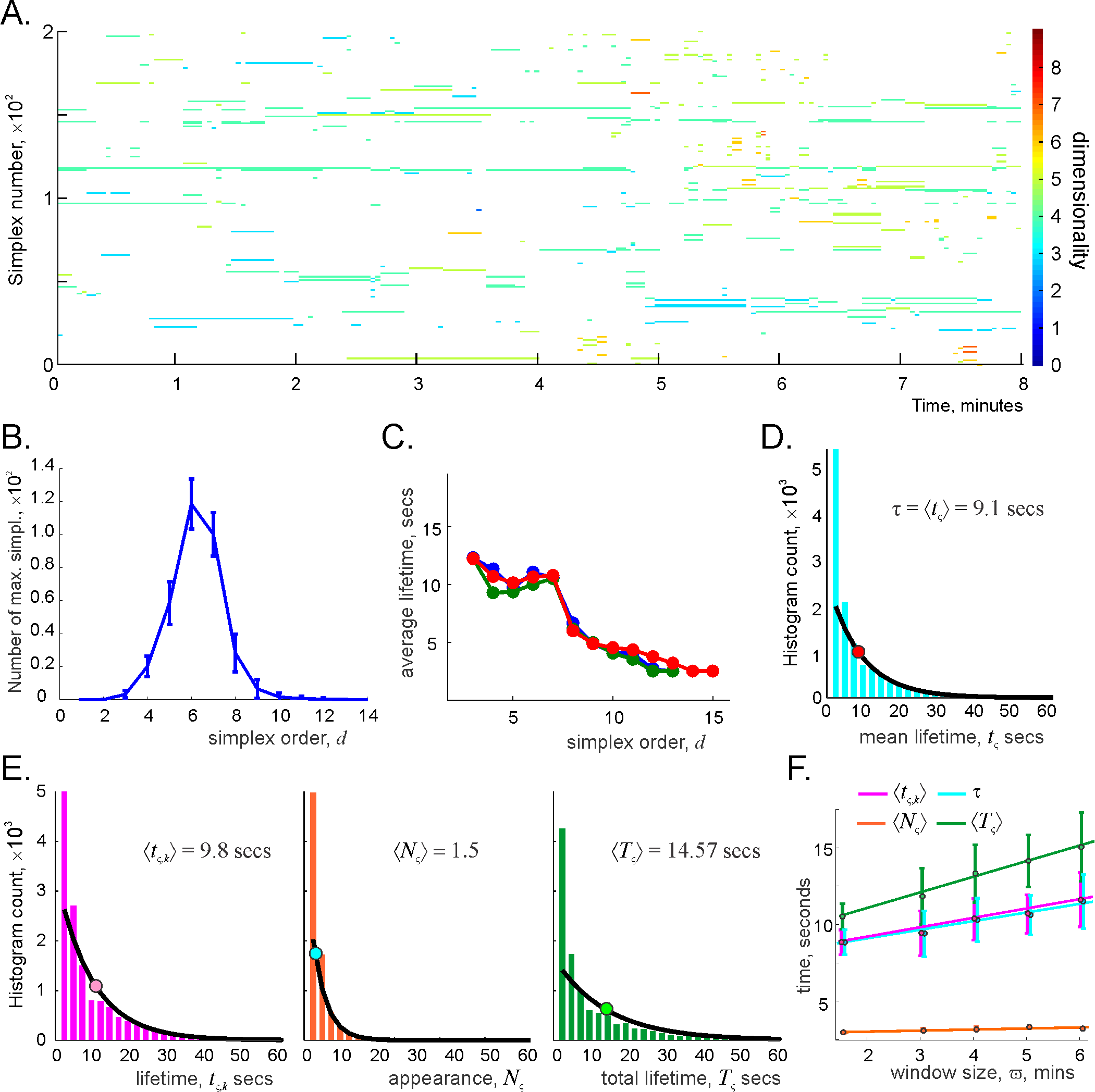}
\caption{\label{Figure4} {\footnotesize \textbf{Fluctuating simplexes}. (\textbf{A}) Maximal simplex timeline diagram: each strike 
represents a timeline of a particular maximal simplex $\varsigma$, computed for the coactivity window $\varpi = 4$ min. 
There are about $N_{\varsigma} = 320$ maximal simplexes at every given timestep (first 200 are enumerated along the 
$y$-axis), whereas the total number of maximal simplexes observed during the entire navigation period is about 11,000. 
The color of the timelines marks the order of $\varsigma$ (colorbar on the right). Notice that the simplexes of 
lower orders generally persist over longer intervals. (\textbf{B}) Number of maximal simplexes as a function of their 
order has a Gaussian shape with the mean $d = 7$ and standard deviation $\Delta_d \approx \pm 2$, suggesting 
that a typical cell assembly contains about seven neurons and about two neurons may appear or disappear in it at a given 
moment. (\textbf{C}) Average existence time of the maximal simplexes tends to decay with increasing order. 
An exception is provided by the lowest order ($1D$) connections, which rarely appear as independently and quickly become 
absorbed into higher order maximal simplexes. (\textbf{D}) Histogram of the maximal simplexes' individual average lifetimes 
$\tau_{\varsigma}$ fits with the exponential distribution with mean $\tau = 9$s, defining the half-life of the simulated cell 
assemblies for this $\varpi$. (\textbf{E}) Histogram of the maximal simplexes' lifetimes $t_{\varsigma,k}$, i.e., histogram of 
the lengths of all intervals between consecutive appearance and disappearance of the maximal simplexes, the histogram of 
the number of simplex-births $N_{\varsigma}$ and the histogram of the total existence periods $T_{\varsigma}$ fit with 
their respective exponential distributions. The mean number of simplex' appearances $\langle N_{\varsigma}\rangle \approx 1.5$ 
shows that most maximal simplexes appear only once or twice, though some maximal simplexes may appear 20 times or 
more. Notice that the mean net existence period $\langle T_{\varsigma} \rangle \approx 14.57$s is approximately equal 
to the product of the mean lifetime and the mean number of appearances 
$\langle T_{\varsigma} \rangle \approx \langle N_{\varsigma}\rangle\langle t_{\varsigma,k} \rangle$. (\textbf{F}) As the 
size of the memory window $\varpi$ increases, the lifetimes, half-lives, and net existence periods of the maximal simplexes 
grow linearly with $\varpi$.}}
\end{figure}

\textbf{Flickering cell assemblies}. We studied the dynamics the flickering cell assemblies produced by a neuronal 
ensemble containing $N_{c} = 300$ simulated place cells. First, we built a simulated cell assembly network (see 
Methods and \cite{Babichev1}) that contains, on average, about $N_{\varsigma} \approx 320$ finite lifetime---transient---cell 
assemblies (Fig.~\ref{Figure4}A). As shown in Fig.~\ref{Figure4}B, the order of the maximal simplexes that represent 
these assemblies ranges between $|\varsigma| = 2$ and $|\varsigma| = 14$, with the mean of about $|\varsigma| = 7$, 
implying that a typical simulated cell assembly includes $|\varsigma| = 7\pm2$ cells. 

The distribution of the maximal simplexes' lifetimes $\delta t_{\varsigma,k}$ as a function of their dimensionality shows 
that higher-dimensional simplexes (and hence the higher-order cell assemblies) are shorter lived than the low order cell 
assemblies (Fig.~\ref{Figure4}C). The histogram of the mean lifetimes $t_{\varsigma} = \langle\delta t_{\varsigma,k}\rangle_{k}$ 
is closely approximated by the exponential distribution (Fig.~\ref{Figure4}D), which suggests that the duration of the cell 
assemblies' existence can be characterized by a half-life $\tau$. The individual lifetimes $\delta t_{\varsigma,k}$, the number 
of appearances $N_{\varsigma}$, and net existence time $\Delta T_{\varsigma} = \sum_{k} \delta t_{\varsigma,k}$ of the 
maximal simplexes and of pairwise connections are also exponentially distributed (see Fig.~\ref{Figure4}E and Fig.~\ref{SFigure1}). 
As expected, the mean net existence time approximately equals to the product of the mean lifetime and the mean number 
of the cell assembly's appearance 
$\langle\Delta T_{\varsigma}\rangle \approx \langle N_{\varsigma}\rangle\langle\delta t_{\varsigma,k}\rangle$. 

Fig.~\ref{Figure4}F shows how these parameters depend on the width of the integration window. As $\varpi$ widens, mean 
lifetime $t_{\varsigma}$ of maximal simplexes (and hence its half-life and the net lifetime) grows linearly, whereas the number 
of appearances $\langle N_{\varsigma}\rangle$ remains nearly unchanged. The latter result is natural since the frequency 
with which the cell assemblies ignite is defined by how frequently the animal visits their respective cell assembly fields, i.e., 
the domains where the corresponding sets of place fields overlap \cite{Babichev1}). This frequency does not change 
significantly if the changes in $\varpi$ do not exceed the characteristic time required to turn around the maze and revisit cell 
assembly fields, in this case ca. $1-2$ min. Thus, the model produces a population of rapidly changing cell assemblies; in the 
simulated case $\tau \approx 9$ seconds, which is close to the experimental range of values \cite{Buzsaki1}. This allows 
us to address our main question: can a network of transient cell assemblies encode the topology of the environment?

\begin{figure}[!hbt] 
\includegraphics[scale=0.84]{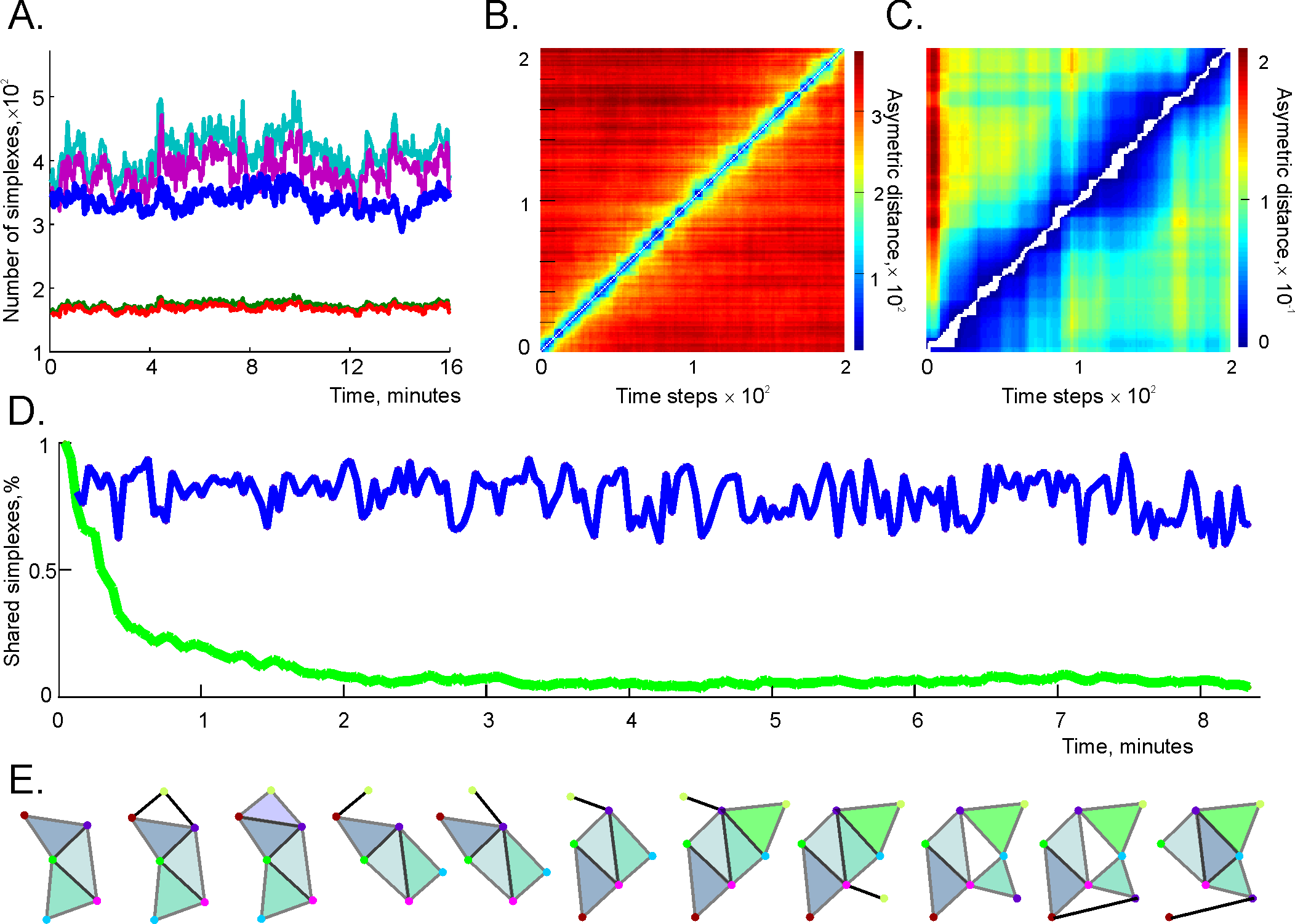}
\caption{\label{Figure5} {\footnotesize \textbf{Behavior of the flickering coactivity complex} computed for the memory window width 
$\varpi = 4$ min, shifted over $\Delta t = 2.5$ secs steps. (\textbf{A}) The number of maximal simplexes in 
$\mathcal{F}_{\varpi}$ (blue trace) fluctuates within $4\%$ of the mean value of $N_{\varsigma} = 320$. The number 
of the $1D$ simplexes $N_{1}(\mathcal{F}_{\varpi})$ (red trace) and the number of the $1D$ simplexes appearing in 
consecutive windows (i.e., links shared by $\mathcal{F}_{\varpi}(t_{i})$ and $\mathcal{F}_{\varpi}(t_{i-1})$, green trace) 
fluctuate within a $3\%$ bound. The fluctuations in the number of $2D$ subsimplexes ($N_{2}(\mathcal{F}_{\varpi})$, 
light blue trace) and the number of $2D$ simplexes shared by two consecutive windows (purple trace) do not exceed 
$7\%$ of the mean. $N_{1}(\mathcal{F}_{\varpi})$ and $N_{2}(\mathcal{F}_{\varpi})$ are scaled down by a factor 
of 10 to fit the scale of the figure. (\textbf{B}) The asymmetric distance between $\mathcal{F}_{\varpi}(t_{i})$ and 
$\mathcal{F}_{\varpi}(t_{j})$ is defined as the number of the maximal simplexes at moment $t_{i}$ which are missing 
at the moment $t_{j}$ for all pairs ($t_i$, $t_i$). As the temporal separation $|t_{i} - t_{j}|$ grows, the asymmetric 
distance between $\mathcal{F}_{\varpi}(t_{i})$ and $\mathcal{F}_{\varpi}(t_{j})$ rapidly increases. (\textbf{C}) The 
matrix of similarity coefficients $r_{ij}$ between the weighted coactivity graphs at different moments of time. For close 
moments $t_{i}$ and $t_{j}$ the differences between $G(t_{i})$ and $G(t_{j})$ are small, but as time separation grows, 
the differences accumulate, though not as rapidly as with the coactivity complexes. (\textbf{D}) At each $t_{i}$ the blue 
line shows the proportion of maximal simplexes present at the previous time, $t_{i-1}$. The green line shows the proportion 
of maximal simplexes contained at the start (in $\mathcal{F}_{\varpi}(t_{1}))$ that remain in the coactivity complex at 
the later time $\mathcal{F}_{\varpi}(t_{i})$. The population of simplexes changes by about $0.95\%$ in about 2 min. 
(\textbf{E}) A schematic illustration of the changes of the coactivity complex's shape: the fluctuations induce permanent 
restructurings. No skeletal structure, similar to that in the ``expected'' scenario shown in Fig.~\ref{Figure2}B, is preserved.}}
\end{figure}

\textbf{Flickering coactivity complex}. We next studied the flickering coactivity complex $\mathcal{F}_{\varpi}$ formed by 
the pool of fluctuating maximal simplexes. First, we observed that the size of $\mathcal{F}_{\varpi}$ does not fluctuate 
significantly across the rats' navigation time. As shown in Fig.~\ref{Figure5}A, the number of maximal simplexes 
$N_{\varsigma}(\mathcal{F}_{\varpi}(t))$ fluctuates within about $4\%$ of its mean value. The fluctuations in the number 
of coactive pairs $N_{2}(\mathcal{F}_{\varpi}(t))$ is even smaller: $3\%$ of the mean, and the variations in number of 
the third order simplexes $N_{3}(\mathcal{F}_{\varpi}(t))$ are about $7\%$ of the mean. To quantify the structural changes
in $\mathcal{F}_{\varpi}$, we computed the number of maximal simplexes that are present at time $t_{i}$ and missing at 
time $t_{j}$, yielding the matrix of asymmetric distances, 
$d_{ij} = N_{\varsigma}(\mathcal{F}_{\varpi}(t_{i})\ \mathcal{F}_{\varpi}(t_{j}))$ for all pairs $t_{i}$ and $t_{j}$ (see 
Methods and Fig.~\ref{Figure5}B). The result suggests that as temporal separation $|t_{i} - t_{j}|$ increases, the differences 
between $\mathcal{F}_{\varpi}(t_{i})$ and $\mathcal{F}_{\varpi}(t_{i})$ rapidly accumulate, meaning that the pool of 
maximal simplexes shared by $\mathcal{F}_{\varpi}(t_{i})$ and $\mathcal{F}_{\varpi}(t_{i})$ rapidly thins out. After 
about $50$ timesteps ($|i - j| > 50$) the difference is about $95\%$ (Fig.~\ref{Figure5}B).

Since the coactivity complexes are induced from the pairwise coactivity graph $G$ as clique complexes, we also studied 
the differences between the coactivity graphs at different moments of time by computing the normalized distance between 
the coactivity matrices (see Methods). The results demonstrate that the differences in $G$, i.e., between $G(t_{j})$ and 
$G(t_{i})$, accumulate more slowly with temporal separation than in $\mathcal{F}_{\varpi}$: after about two minutes the 
connectivity matrices differ by about $10-15\%$ (Fig.~\ref{Figure5}C).

The Fig.~\ref{Figure5}D shows the asymmetric distance between two consecutive coactivity complexes 
$\mathcal{F}_{\varpi}(t_{i})$ and $\mathcal{F}_{\varpi}(t_{i+1})$, and the asymmetric distance between the starting 
and a later point $\mathcal{F}_{\varpi}(t_{1})$ and $\mathcal{F}_{\varpi}(t_{i})$, normalized by the size of 
$\mathcal{F}_{\varpi}(t_{1})$ as a function of time. The results suggest that, although the sizes the coactivity complexes 
at consecutive time steps do not change significantly, the pool of the maximal simplexes in $\mathcal{F}_{\varpi}$ is nearly 
fully renewed after about two minutes. In other words, although the coactivity complex changes its shape slowly, the 
integrated changes across long periods are significant (compare Fig.~\ref{Figure5}E with Fig.~\ref{Figure2}B). Biologically, 
this implies that the simulated cell assembly network, as described by the model, completely rewires in a matter of minutes.            

\begin{figure} 
\includegraphics[scale=0.84]{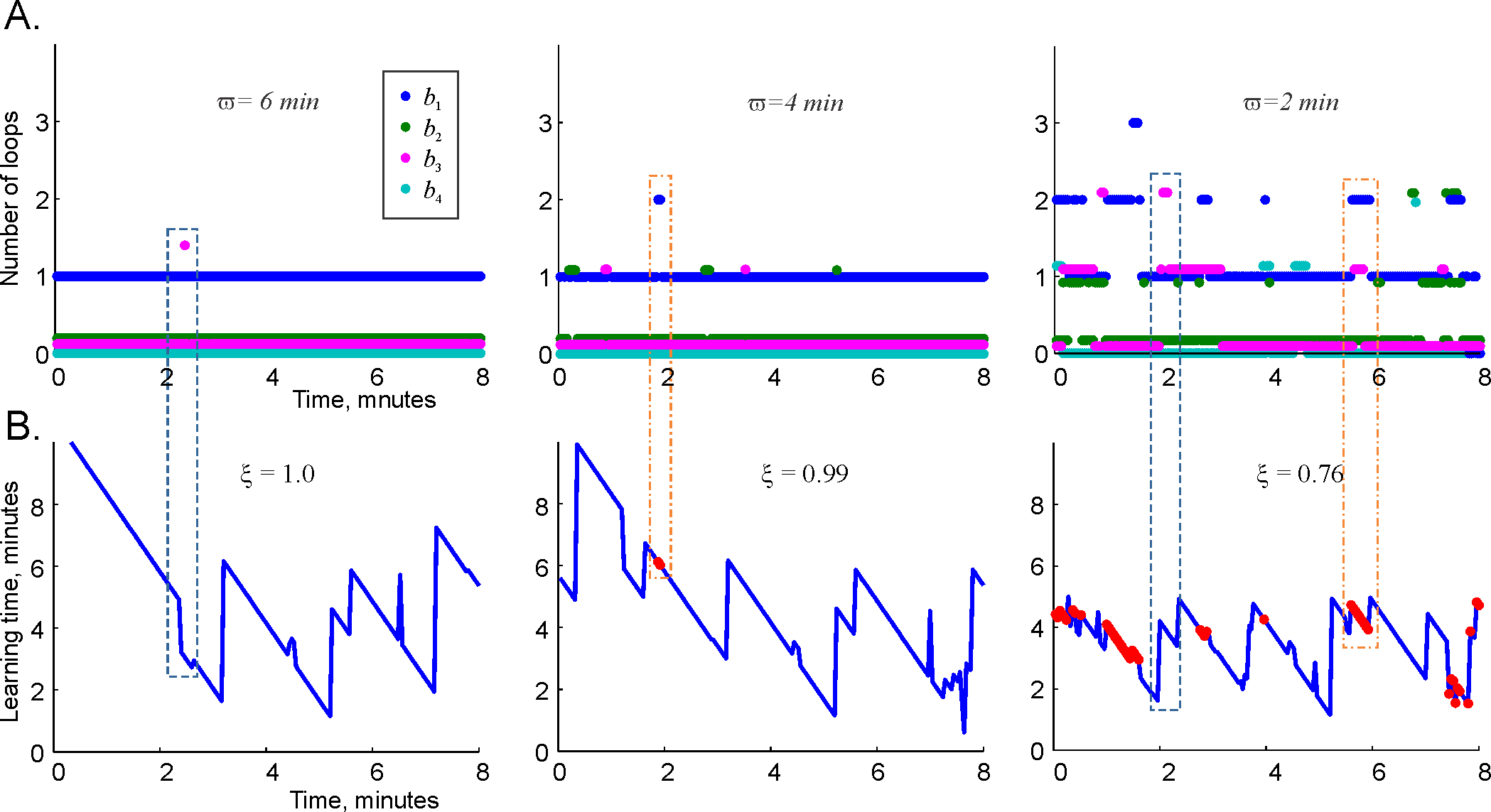}
\caption{\label{Figure6} {\footnotesize \textbf{Stability of large-scale topological information}. (\textbf{A}) The low-dimensional Betti 
numbers $b_{1}$, $b_{2}$, $b_{3}$ as a function of discrete time, computed for three coactivity integration windows, 
$\varpi = 2$ min, $\varpi = 4$ min, and $\varpi = 6$ min. The $0$-th Betti number, $b_{0} = 1$, remains stable at all 
times and is therefore not shown. At sufficiently large coactivity windows, $\varpi \sim 4-6$ minutes, the topological 
fluctuations become suppressed and the large-scale topological information remains stable, even though the characteristic 
lifetime of a maximal simplex in the coactivity complex $\mathcal{F}_{\varpi}$ is about $10$ secs (Fig.~\ref{Figure4}C). 
As the integration window narrows, the topological fluctuations intensify (Fig.~\ref{SFigure2}). (\textbf{B}) The 
variation in the time required to extract the topological information increases as the coactivity integration window narrows. 
At $\varpi = 2$ min, the complex $\mathcal{F}_{\varpi}$ fails to produce the correct topological information in $24\%$ 
of the cases (convergence score $\xi = 0.76$). The failing moments are marked by red dots. As the memory window 
increases to $\varpi = 4$ min, the topological mapping fails in only $1\%$ of the cases. As the memory window increases 
to $\varpi = 6$ min, the failing points disappear. Here we compute the most conservative estimate for the learning time 
$T_{\min}$ to be the time required to establish the correct topology only in the dimensions that may contain physical 
obstructions, $0D$ and $1D$. Therefore, the points where $T_{\min}$ diverges are marked by appearances of spurious 
$1D$ loops (encapsulated into red dashed boxes across panels). The points where the learning time rapidly changes are 
often accompanied by the appearance or disappearance of higher dimensional topological loops (blue dashed boxes).}}
\end{figure}

\textbf{Topological analysis of the flickering coactivity complex} exhibits a host of different behaviors. First, we start by 
noticing that the 0th and the higher-order Betti numbers always assume their physical values $b_{0} = 0$, $b_{n > 4} = 0$, 
whereas the intermediate Betti numbers $b_{1}$, $b_{2}$, $b_{3}$ and (for small $\varpi$s) $b_4$ may fluctuate 
(Fig.~\ref{Figure6}A and Fig.~\ref{SFigure2}). Thus, despite the fluctuations of its simplexes, the flickering complex 
$\mathcal{F}_{\varpi}$ does not disintegrate into pieces and remains contractible in higher dimensions ($D > 3$). 
Biologically, this implies that the topological fluctuations in the simulated hippocampal map are limited to $1D$ loops, $2D$ 
surfaces and $3D$ bubbles. For example, an occurrence of $b_{1} = 2$ value indicates the appearance of an extra (non-physical) 
$1D$ loop that surrounds a spurious gap in the cognitive map (Fig.~\ref{Figure1}A). On the other hand, at the moments 
when $b_{1} = 0$, all $1D$ loops in $\mathcal{F}_{\varpi}$ are contractible, i.e., the central hole is not represented in 
the simulated hippocampal map. The moments when $b_{n >2} > 0$ indicate times when the flickering complex 
$\mathcal{F}_{\varpi}$ contains non-physical, non-contractible multidimensional topological surfaces. One can speculate 
about the biological implications of these fluctuations, see Fig.~\ref{SFigure4}.
 
As the coactivity window increases, the fluctuating topological loops become suppressed and vice versa. As the integration 
window shrinks, the fluctuations of the topological loops intensify (Fig.~\ref{Figure6}). This tendency could be expected, 
since the cell assembly lifetimes reduce as the integration window shrinks and increase as the coactivity integration 
window grows (Fig.~\ref{Figure4}F). However, a nontrivial result suggested by Fig.~\ref{Figure6} is that the topological 
parameters of the flickering complex can stabilize completely, even though its maximal simplexes keep appearing and 
disappearing, or ``flickering.'' At $\varpi \approx 6$ minutes, the Betti numbers of $\mathcal{F}_{\varpi}$ remain 
unchanged (Fig.~\ref{Figure6}A), whereas the lifetime of its typical simplex is about 10 seconds (Fig.~\ref{Figure4}F). 
Biologically, this implies that a stable hippocampal map can be encoded by a network of transient cell assemblies, i.e., that 
the ongoing synaptic plasticity in the hippocampal network does not necessarily compromise the integrity of the large-scale 
representation of the environment. 

\textbf{Local learning times}. If the information about the detected place cell coactivities is retained indefinitely, the time 
required for producing the correct topological barcode of the environment $T_{\min}$ may be computed only once, starting 
from the onset of the navigation, and used as the low-bound estimate for the learning time \cite{PLoS,Arai}. In the case of 
a rewiring (transient) cell assembly network, the pool of encoded spatial connectivity relationships is constantly renewed. 
As a result, the time required to extract the large-scale topological signatures of the environment from place cell coactivity 
becomes time-dependent and its physiological interpretation also changes. $T_{\min}(t_{k})$ now defines the period over 
which the topological information emerges from the ongoing spiking activity at every stage of the navigation.

As shown in Fig.~\ref{Figure6}B, the proportion $\xi$ of ``successful'' coactivity integration windows, i.e., those windows 
in which $T_{\min}$ assumes a finite value, depends on their width $\varpi$. For small $\varpi$, the coactivity complex 
frequently fails to reproduce the topology of the environment (Fig.~\ref{Figure6}A). As $\varpi$ grows, the number of failing 
points, i.e., those for which $T_{\min}(t_{k}) > \varpi$, reduces due to the suppression of topological fluctuations. 
The domains previously populated by the divergent points are substituted with the domains of relatively high but still finite 
$T_{\min}(t_{k})$. For sufficiently large coactivity windows ($\varpi > 6$ minutes), such divergent points become exceptional: 
the correct topological information exists at all times, even though period required to produce this information is time-dependent. 

The time dependence of $T_{\min}(t_{k})$ exhibits abrupt rises and declines, with characteristic $45^{\circ}$ slants in-between. 
The rapid rises of $T_{\min}(t_{k})$ correspond to appearances of obstructions in the coactivity complex 
$\mathcal{F}_{\varpi}$ (and possibly higher-dimensional surfaces) that temporarily prevent certain spurious loops from 
contracting. As more connectivity information is supplied by the ongoing spiking activity, the coactivity complex 
$\mathcal{F}_{\varpi}$ may acquire a combination of simplexes that eliminates these obstructions, allowing the unwanted 
loops to contract and yielding the correct topological barcode. Thus, Fig.~\ref{Figure6}B suggests that the dynamics of the 
coactivity complex is controlled by a sequence of coactivity events that produce or eliminate topological loops in 
$\mathcal{F}_{\varpi}$, while the $45^{\circ}$ slants in $T_{\min}(t_{k})$ represent ``waiting periods'' between these 
events (since with each window shift over $\Delta t$, the local learning time decreases by exactly the same amount).

\begin{figure} 
\includegraphics[scale=0.84]{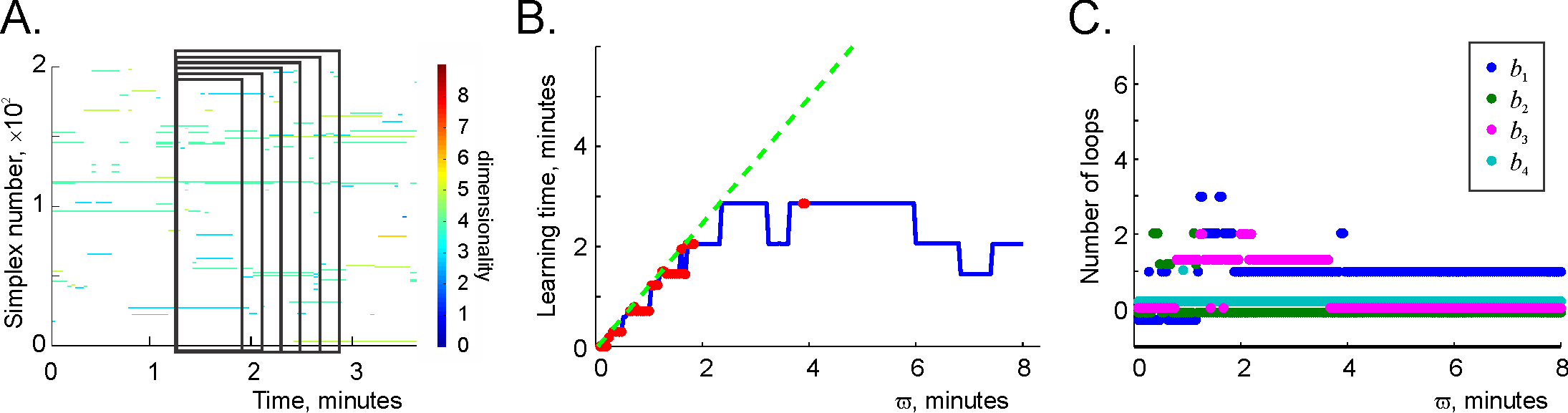}
\caption{\label{Figure7} {\footnotesize \textbf{Stability the large-scale topological information}. (\textbf{A}) A schematic 
illustration of the growing coactivity window $\varpi$, superimposed over a fragment of the maximal simplex timeline diagram. 
(\textbf{B}) The learning times $T_{\min}(\varpi_{k})$ computed within the growing coactivity window. The learning 
times computed within narrow coactivity windows either diverge ($T_{\min}(\varpi_{k}) > \varpi$) or converge barely 
($T_{\min} (\varpi_{k}) \approx \varpi$). As $\varpi$ exceeds a certain critical value $\varpi_{c}$ (for the simulated 
place cell ensemble, $\varpi_{c} \approx 4-6$ minutes), the learning time $T_{\min}(\varpi_{k})$ stops increasing and 
begins to fluctuate around a certain mean value $T_{\min} = \langle T_{\min} (\varpi_{k})\rangle_k$. This value is 
independent of the coactivity window width and hence represents a parameter-free characterization of the mean time 
required to extract topological information from place cell coactivity. (\textbf{C}) The low-dimensional Betti numbers 
$b_{1}$, $b_{2}$, $b_{3}$ and $b_{4}$ as a function of the coactivity integration window width $\varpi$. As $\varpi$ 
exceeds a critical value $\varpi_{c}$, the Betti numbers $b_{n}$ stabilize, indicating suppression of the topological 
fluctuations in $\mathcal{F}_{\varpi}$.}}
\end{figure} 

It should be noticed that the network's failure to produce a topological barcode at a particular moment, i.e., within a 
particular integration window $\varpi_{k}$, is typically followed by a period of successful learning. This implies that the 
rudimentary forgetting mechanism incorporated into the model, whereby the removal of older connectivity relationships 
from $\mathcal{F}_{\varpi}$ as newer relationships are acquired, allows correcting some of the accidental connections 
that that may have been responsible for producing persistent spurious loops at previous steps. In other words, a network 
capable of not only accumulating, but also forgetting information, exhibits better learning results.

Thus, the process of extracting the large-scale topology of the environment should be quantified in terms of the mean 
learning time $T_{\min}  = \langle T_{\min}(t_{k})\rangle_{k}$ and its variance $\Delta T_{\min} /T_{\min}$, which 
does not exceed $40\%$ (typically $\Delta T_{\min} /T_{\min} \approx 20\%$). This suggests that $T_{\min}$ provides 
a statistically sound characteristics of the information flow across the simulated cell assembly network.

To better understand how the learning time depends on the memory width, we tested the dependence of $T_{\min}$ on the 
size of the coactivity integration window $\varpi$. We fixed the position of several coactivity integration windows $\varpi_{k}$ 
and expanded their right side, $\varpi_{k}^{(1)} > \varpi_{k}^{(2)} > ... > \varpi_{k}^{(q)}$ (Fig.~\ref{Figure7} and 
Fig.~\ref{SFigure3}). As one would expect, small values of $\varpi$ generated many failing points, whereas the learning times 
$T_{\min}(t_{k})$ computed for the successful trials remained nearly equal to $\varpi$, i.e., the width of the narrow integration 
windows was barely sufficient for producing the correct barcode $\mathfrak{b}(\mathcal{E})$. However, as $\varpi$ grows 
further, $T_{\min}$ stops increasing and, as $\varpi$ exceeds a certain critical value $\varpi_{c}$ (typically about five or six 
minutes), the learning time begins to fluctuate around a mean value $T_{\min} = \langle T_{\min} (t_{k})\rangle$ of about 
two minutes. In other words, for sufficiently large coactivity windows $\varpi > \varpi_{c}$, the learning times become 
\emph{independent} of the model parameter $\varpi$ and therefore provides a parameter-free characterization of the time 
required by a network of place cell assemblies to represent the topology of the environment, whereas $\varpi_c$ defines the 
time necessary to collect the required spiking information (Fig.~\ref{SFigure4}).

\section{Discussion}
\label{section:discussion}

Fundamentally, the mechanism of producing the hippocampal map depends on two key constituents: on the timing of the 
action potentials produced by the place cells and by the way in which the spiking information is processed by the downstream 
networks. A key determinant for the latter is the synaptic architecture of the cell assembly network, which changes constantly 
due to various forms of synaptic and structural plasticity: place cell assemblies may emerge in cell groups that exhibit frequent 
coactivity or disband due to lack thereof. The latter phenomenon is particularly significant: since the hippocampal network is 
believed to be one of the principal memory substrates, frequent recycling of synaptic connections may compromise the integrity 
of its net function. For example, the existence of many-to-one projections from the CA3 to the CA1 region of the hippocampus 
suggests that the CA1 cells may serve as the readout neurons for the assemblies formed by the CA3 place cells \cite{Carr,Johnson}. 
Electrophysiological studies suggest that the recurrent connections within CA3 and the CA3-CA1 connections rapidly renew during 
the learning process and subsequent navigation \cite{Sasaki,Keller}. On the other hand, it is also well known that lesioning these 
connections disrupts the animal's performance in spatial \cite{Lee1,Kim,Kesner1} and nonspatial \cite{Eichenbaum1,Farovik} 
learning tasks, which suggests that an exceedingly rapid recycling of functional cell groups may impair the net outcome of the 
hippocampal network, which is the hippocampal spatial map \cite{Madronal,Gilbert,Lee2,Steffenach}.

The proposed model allows investigating whether a plastic, dynamically rewiring network of place cell assemblies can sustain 
a stable topological representation of the environment. The results suggest that if the intervals between consecutive appearance 
and disappearance of the cell assemblies are short, the hippocampal map exhibits strong topological fluctuations. However, if 
the cell assemblies rewire sufficiently slowly, the information encoded in the hippocampal map remains stable despite the 
connectivity transience in its neuronal substrate. Thus, the plasticity of neuronal connections, which is ultimately responsible 
for the network's ability to incorporate new information \cite{McHugh,Leuner, Dupret,Schaefers}, does not necessarily degrade 
the information that is already stored in the network. These results present a principal development of the model outlined in 
\cite{PLoS,Arai,Babichev1} from both a computational and a biological perspective.

\textbf{Physiological vs. schematic learnings}. The schematic approach proposed in \cite{Schemas} allows describing 
the process of spatial learning from two perspectives: as training of the synaptic connections within the cell assembly 
network---referred to as physiological learning in \cite{Schemas}---or as the process of establishing large-scale topological 
characteristics of the environment, referred to as ``schematic,'' or ``cognitive,'' learning. The difference between these 
two concepts is particularly apparent in the case of the rewiring cell assembly network, in which the synaptic configurations 
may remain unsettled due to the rapid transience of the connections. On the other hand, schematic learning is perfectly 
well defined since the large-scale topological characteristics of the environment can be achieved reliably.

In fact, the model outlines three spatial information processing dynamics at the short-term, intermediate-term, and long-term 
memory timescales \cite{Cowan}. First, local spatial connectivity information is represented in transient cell assemblies within 
several seconds. This timescale corresponds to the scope of memory processes that involve temporary maintenance of information 
produced by the ongoing neural spiking activity, commonly associated with short-term memory \cite{Cowan,Hebb}. The short-term 
memory capacity is around seven ($7 \pm 2$, \cite{Miller}) items, corresponding in the model to the order of the 
simulated cell assemblies (Fig.~\ref{Figure4}B). The information about the large-scale connectivity of the environment is 
acquired and updated---the (mean) learning time $T_{\min}$, Figs.~\ref{Figure6} and \ref{Figure7}---is on the order of 
minutes, corresponding to intermediate-term memory timescale \cite{Eichenbaum2,Kesner2}. The persistent topological 
information, represented by the stable Betti numbers, may represent long-term memory.

\section{Methods}
\label{section:methods}

\textbf{The rat's movements} were modeled in a small planar environment, similar to the arenas used in electrophysiological 
experiments (bottom of Fig.~\ref{Figure1}A). The trajectory covers the environment uniformly, without artificial favoring of 
one segment of the environment over another. 

\textbf{Place cell spiking activity} is modeled as a stationary temporal Poisson process with a spatially localized Gaussian 
rate characterized by the peak firing amplitude $f_c$ and the place field size $s_c$ \cite{Barbieri}. In the simulated ensemble 
of $N_{c} = 300$ place cells, the peak firing amplitudes are log-normally distributed with the mode $f_c = 14$ Hz and the 
place field sizes are log-normally distributed with the mode $s_c = 17$ cm. The place cell spiking probability is modulated 
by the $\theta$-component of the extracellular field oscillations (mean frequency of $\sim 8$ Hz \cite{Buzsaki2}) recorded 
in wild-type Long Evans rats (see Methods in \cite{eLife}). For more computational details see Methods in \cite{PLoS,Arai}.

\textbf{The activity vector} of a place cell $c$ is constructed by binning its spike trains into an array of consecutive coactivity 
detection periods $w$. If the time interval $T$ splits into $N_w$ such periods, then the activity vector of a cell $c$ over this 
period is $m_c(T) =[m_{c;1}, ..., m_{c;N_{w}}]$, where $m_{c;k}$ specifies how many spikes were fired by $c$ into the 
$k$-th time bin \cite{Babichev1}. The activity vectors of $N_{c}$ cells, combined as rows of a $N_{c}\times N_{w}$ matrix, 
form the \emph{activity raster} $R$. A \emph{binary raster} $B$ is obtained from the activity raster $R$ by replacing the 
nonzero elements of $R$ with 1.

\textbf{Place cell spiking coactivity} is defined as the firing that occurred over two consecutive $\theta$-cycles, which is an 
optimal coactivity detection period $w$ both from the computational \cite{Arai} and from the physiological \cite{Mizuseki} 
perspective. A coactivity $\rho$ of a pair of cells $c_{1}$ and $c_{2}$ can be computed as the formal dot product of their 
respective activity vectors $\rho_{c_{1}c_{2}} = m_{c_{1}}(T) m_{c_{2}}(T)$.

\textbf{Shifting coactivity window}. The spiking activity confined within the $k$-th coactivity integration window of size 
$\varpi$ produces a local binary raster $B_{k}$ of size $N_{c} \times N_{\varpi/w}$, where $N_{\varpi/w} = \lfloor\varpi /w \rfloor$. 
The coactivity integration window was shifted by the discrete time steps $\Delta t = 10w \approx 2.5$s. 
Thus, in $n_s = \varpi/\Delta t$ steps, the local rasters $B_{k}$ and $B_{k+ns}$ cease to overlap during the four-minute-long 
coactivity integration window $n_s = 96$.

\textbf{Coactivity distances}. For each window $\varpi_n$, we compute the coactivities of every pair of cells 
$$\rho^{n}_{ij} = \sum_{k} B^{n}_{i_{k}} B^{n}_{j_{k}},$$ 
where $B^{n}_{i_{k}}$ is the ``local'' binary raster of coactivities produced within that window. To compare different 
local rasters, we compute the similarity coefficients between them
$$r_{mn} = \sum_{i.j} |\rho^{n}_{ij} - \rho^{m}_{ij}|/\sum_{i.j} |\rho^{n}_{ij} |,$$ 
where indexes $i$,$j$ run over all the cells in the ensemble, illustrated in Fig.~\ref{Figure4}C.

\textbf{The cell assemblies} were constructed within each memory window using the Method II of \cite{Babichev1}, which 
is computationally more stable, produces less maximal simplexes and yields correct vertex statistics for the simulated 
hippocampal network.

\textbf{Topological analyses} were implemented using the JPlex package \cite{JPlex}.

\section{Acknowledgments}
\label{section:acknow}

We thank R. Phenix for his critical reading of the manuscript.
The work was supported by the NSF 1422438 grant and by the Houston Bioinformatics Endowment Fund.

\newpage
\beginsupplement

\section{Supplementary Figures}
\label{section:SupplFigs}

\begin{figure}[ht] 
\includegraphics[scale=0.84]{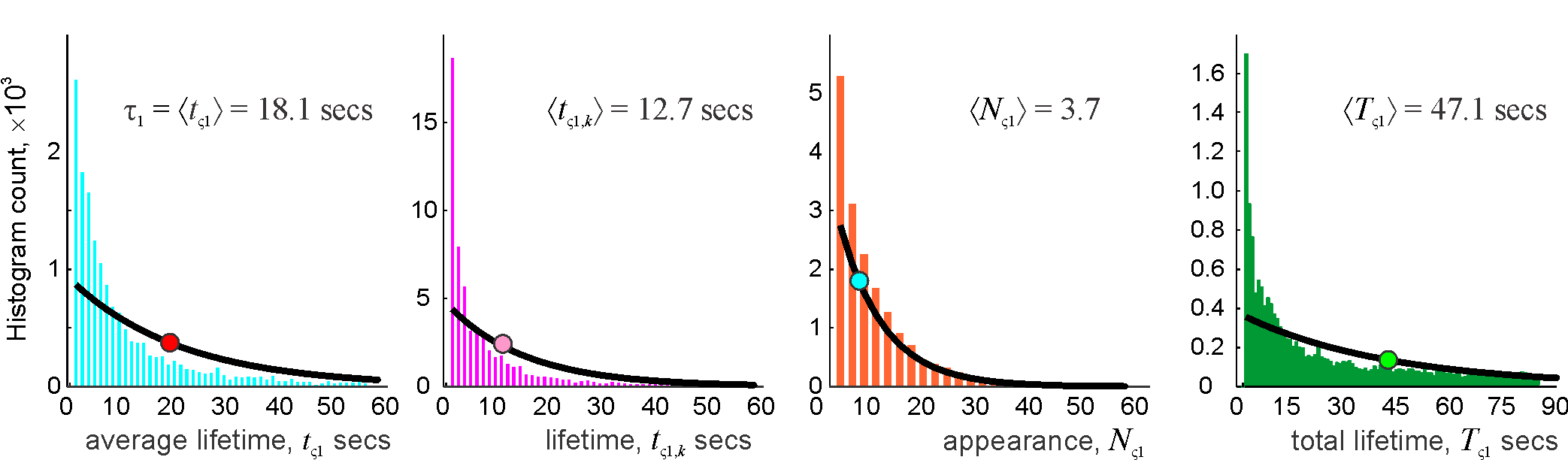}
\caption{\label{SFigure1} {\footnotesize \textbf{The statistics for the flickering connections in the cells assemblies}. The statistics of 
the mean lifetimes $t_{\varsigma_1}$, individual lifetimes $t_{\varsigma_1,k}$, number of appearances $N_{\varsigma_1}$ 
and the total existence periods $T_{\varsigma_1}$ of the $1D$ subsimplexes of the coactivity complex $\mathcal{F}_{\varpi}$, 
which is the links of the coactivity graphs $G(\varpi_n)$, representing connections in the simulated cell assemblies. In all cases, 
the mean net existence period equals approximately to the product of the mean lifetime by the mean number of appearances 
$\langle T_{\varsigma_1}\rangle \approx \langle N_{\varsigma_1}\rangle\langle t_{\varsigma_1,k}\rangle$.}}
\end{figure} 

\begin{figure}[ht] 
\includegraphics[scale=0.84]{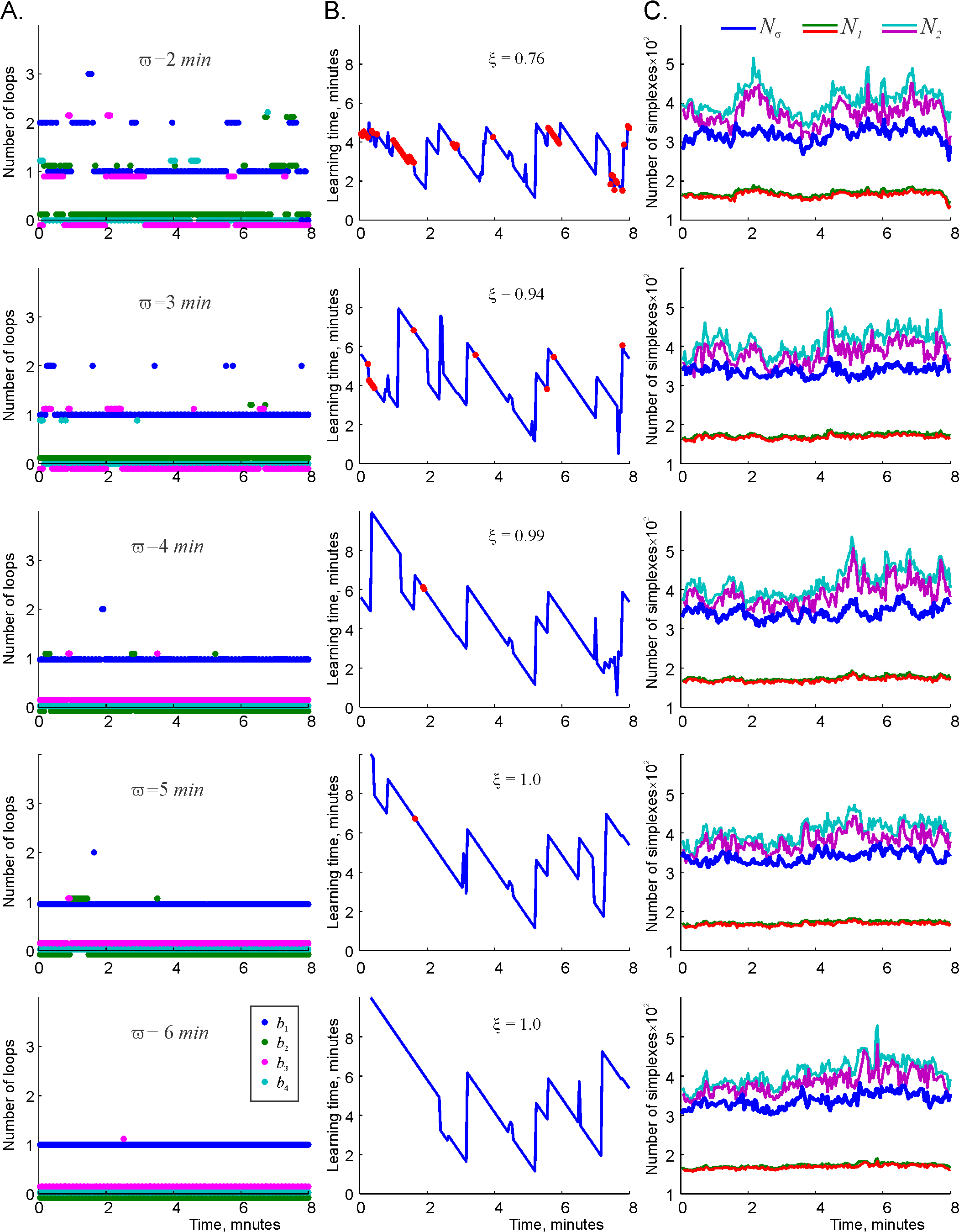}
\caption{\label{SFigure2} {\footnotesize \textbf{Flickering coactivity complex as a function of time}. (\textbf{A}) As the coactivity 
integration window $\varpi$ increases, the topological fluctuations in the coactivity complex $\mathcal{F}_{\varpi}$ are 
suppressed. (\textbf{B}) The corresponding learning times $T_{\min}$. Red dots mark the moments when the map acquires 
a non-physical topological barcode. As the coactivity window $\varpi$ grows, the topological fluctuations are suppressed 
and the number of failures decreases. Notice that the learning time remains high immediately after the areas as the 
failures are suppressed. At $\varpi \approx 5$ min, when the mean half-life of a simulated cell assembly is about 
$\tau_{\varsigma} \approx 10$ secs (Fig.~\ref{Figure4}F), the map retains a topologically correct shape at all times. 
(\textbf{C}) Variations in the size of the coactivity complex $\mathcal{F}_{\varpi}$ reduce with increasing $\varpi$.}}
\end{figure} 

\begin{figure}[ht] 
\includegraphics[scale=0.9]{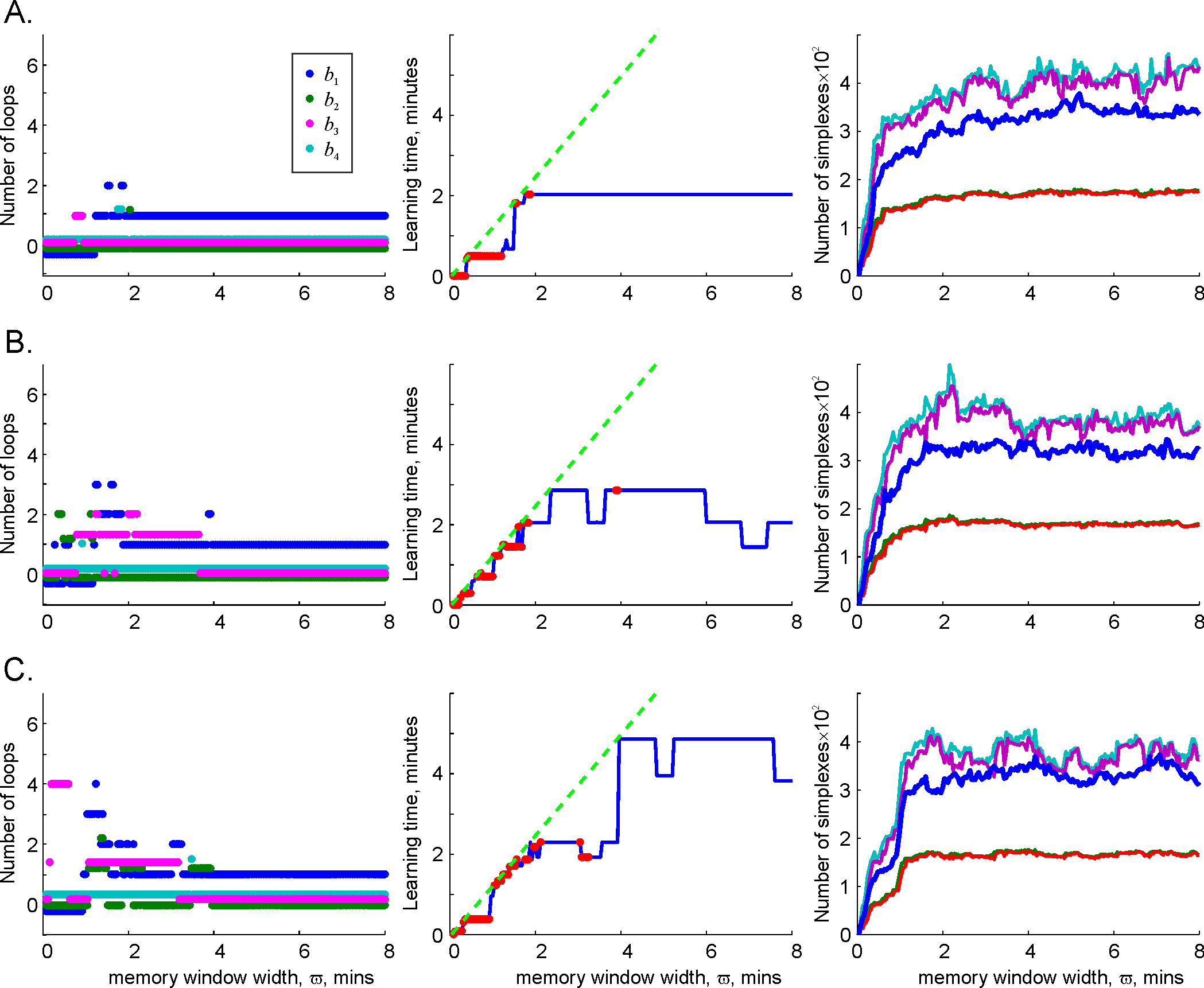}
\caption{\label{SFigure3} {\footnotesize \textbf{Growing memory window}. (\textbf{A}) If the coactivity window is placed at a 
timepoint with few topological fluctuations, the Betti numbers $b_{1}$, $b_{2}$, $b_{3}$ and $b_{4}$, and the learning 
time quickly stabilize. The last panel indicates that size of the coactivity complex $\mathcal{F}_{\varpi}$ grows as a 
function of $\varpi$ and then acquires a stable size. (\textbf{B}) At a typical temporal domain, the behavior of the 
``asymptotic'' coactivity complex $\mathcal{F}_{\varpi}$ exhibits stronger topological fluctuations and the learning 
time fluctuates as a function of increasing $\varpi$. (\textbf{C}) In the locations in which the topological fluctuations 
are strong, the Betti numbers of the flickering coactivity complex $\mathcal{F}_{\varpi}$ take longer to stabilize and 
the learning time may retain high values for longer periods, before returning to the typical regime shown on panel B.}}
\end{figure} 

\begin{figure}[ht] 
\includegraphics[scale=0.9]{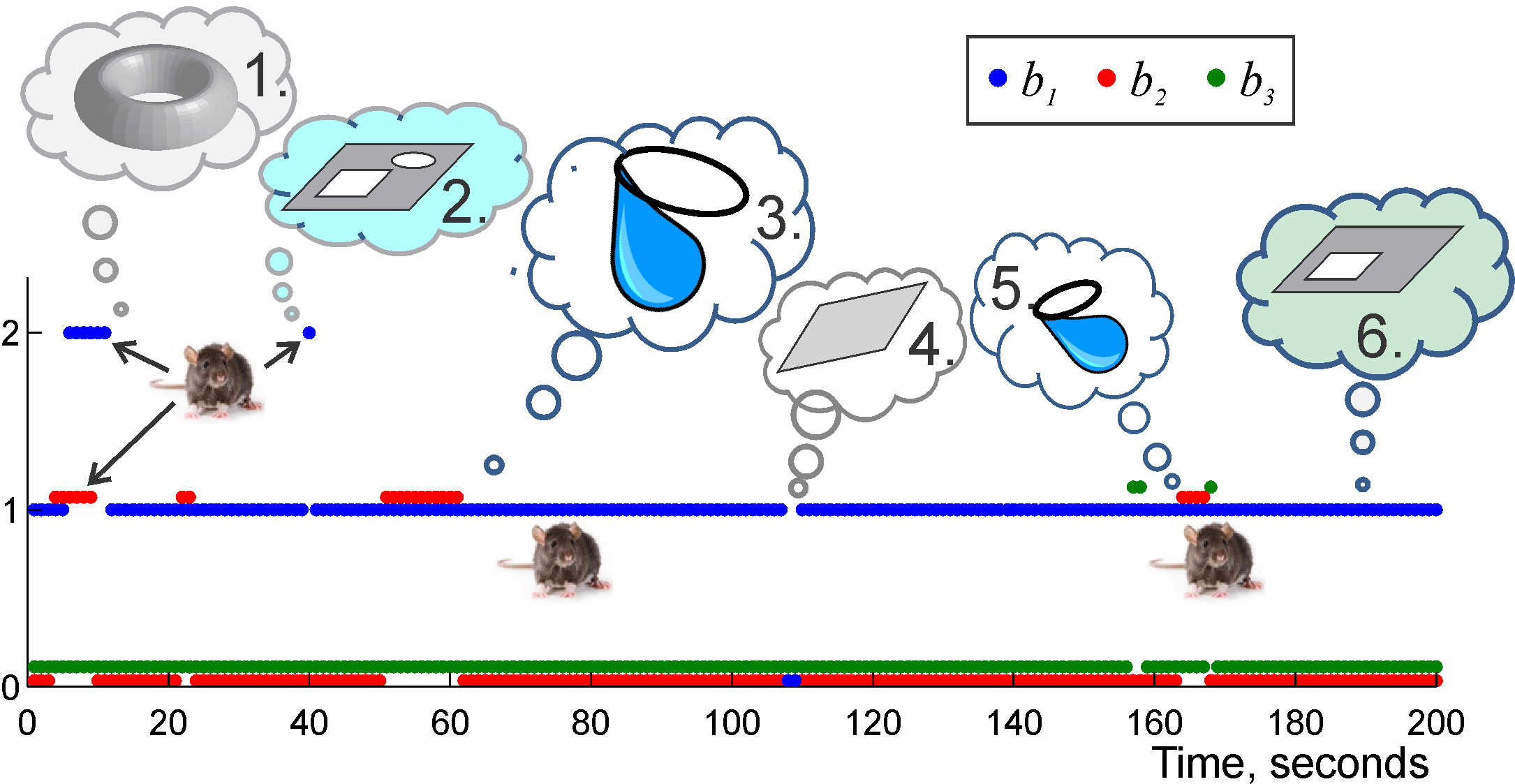}
\caption{\label{SFigure4} {\footnotesize \textbf{Topological fluctuations in the hippocampal map}. As in the case illustrated 
in Fig.~\ref{Figure6}, we assume that the map forms one single piece at all times, hence its $0$-th Betti number 
$b_{0} = 1$ is not shown. At the moment when the Betti numbers assume values $b_{1} = 2$, $b_{2} = 1$ and 
$b_{n >2} = 0$, the hippocampal map warps into a shape that is topologically equivalent to a torus (Fig.~\ref{Figure3}C). 
At the moment when $b_{1} = 2$ while $b_{n >1} = 0$, the map contains an extra $1D$ loop indicating an extra 
gap in $\mathcal{F}_{\varpi}$. At the times when $b_{1} = b_{2} = 1$ while $b_{n >2} = 0$, the map contains a 
$2D$ bulge and a non-contractible cycle fused together (since $b_{0} = 1$!). For most times, the topological type 
of $\mathcal{F}_{\varpi}$ coincides with the topological type of the simulated rat's environment, $b_{1} = 1$, 
$b_{n >1} = 0$.}}
\end{figure} 

\end{document}